\documentclass[aip,cha,reprint]{revtex4-1}

\usepackage{graphicx}
\usepackage{amsmath,amssymb}
\usepackage{hyperref}
\usepackage{rotating}
\usepackage{multirow}

\begin{document}


\title{Contour polygonal approximation using shortest path in networks}

\author{Andr\'{e} Ricardo Backes}
\affiliation{
	     Faculdade de Computa\c{c}\~{a}o - Universidade Federal de Uberl\^{a}ndia \\
             Tel.: +55 34 3239-4144 
             Fax: +55 34 3239-4392 
             email: arbackes@yahoo.com.br } 
\author{Dalcimar Casanova}
\affiliation{
              S\~{a}o Carlos Institute of Physics (IFSC) - University of S\~{a}o Paulo\\
              phone: +55 16 3373 8728
              Fax.: +55 16 3373 9879
              email: dalcimar@ursa.ifsc.usp.br bruno@ifsc.usp.br\\
              http://scg.ifsc.usp.br}  
\author{Odemir Martinez Bruno}
\affiliation{
              S\~{a}o Carlos Institute of Physics (IFSC) - University of S\~{a}o Paulo\\
              phone: +55 16 3373 8728
              Fax.: +55 16 3373 9879
              email: dalcimar@ursa.ifsc.usp.br bruno@ifsc.usp.br\\
              http://scg.ifsc.usp.br}               

\hyphenation{geo-de-sic}

\begin{abstract}
Contour polygonal approximation is a simplified representation of a contour by line segments, so that the main characteristics of the contour remain in a small number of line segments. This paper presents a novel method for polygonal approximation based on the Complex Networks theory. We convert each point of the contour into a vertex, so that we model a regular network. Then we transform this network into a Small-World Complex Network by applying some transformations over its edges. By analyzing of network properties, especially the geodesic path, we compute the polygonal approximation. The paper presents the main characteristics of the method, as well as its functionality. We evaluate the proposed method using benchmark contours, and compare its results with other polygonal approximation methods.
\end{abstract}


\maketitle

\section{Introduction}
\label{sec:introduction}

A boundary curve performs an important role in shape analysis and representation. It is one of the most important visual attributes in image processing and computer vision. A large amount of information of the contour is redundant and it can be simplified without damaging the representation of the original shape. In biological vision systems, this evidence was demonstrated by Atteneave \cite{Attneave-1954}. He did experiments in psychophysics observing whether man would be capable of recognising shapes containing only part of their outline. The experiments of Atteneave demonstrated that points of high curvature observed in objects consist of the most important information about shape, and that the human visual system is capable of recognising the original shape by just observing the higher curvature segments of the contour. This evidence motivated the development of methods to reduce the information of the contour, without losing information from the original shape.

Among these methods of contour curves reduction, there are algorithms for polygonal approximation. These methods preserve the main characteristics of the curve of a contour using as few line segments as possible. The polygonal approximation presents various advantages over other curve representation schemes (various techniques can be seen in \cite{bb16963}) due to its computational simplicity, coding efficient, preservation of local properties and the low complexity of feature extraction. Moreover, an important aspect of curve polygonal approximation is the preservation of shape perception.

Shapes can be represented easily by a set of straight line segments that interconnect high curvature points of an outline. Usually, a human observer can recognise this type of representation. This fact has motivated researches in high curvature points detection and their polygonal approximation.

In this paper, we present a novel method for polygonal approximation of digital curves. The authors have been worked in this problem using the Complex Networks theory. In a previous work \cite{journals/isci/BackesB13} we proposed an algorithm based on vertex betweenness. Here, we focus on the study of the evolution of its shortest path as we convert a regular network into a Small-World network. The vertex betweenness and the shortest path, using different networks properties and characteristics. Although, these approaches are different, they achieved similar general performance. As can be noticed, there are special situations of the problem in which one or other approach is more appropriate.

We present experiments with contours and their respective results. We also compared the method with other polygonal approximation methods. Results demonstrate the great ability of the complex networks techniques to remove redundant point of a digital curve, thus contributing to the problem of the polygonal approximation.

\section{Proposed Method}
\label{sec:method}

In this section, we present the definition of the polygonal approximation problem, followed by the description of how we model the problem as a Complex Network (here referred to as SP).

\subsection{Problem definition}
\label{sec:problem}

Consider $S = \left\{ x_{1},x_{2},...,x_{n}\right\}$ a clockwise ordered set of points describing a curve or object contour. The aim of polygonal approximation is to find out $S^{*} = \left\{ x_{v(1)},x_{v(2)},...,x_{v(m)} \right\}$, with $v(i) \in [1,n]$ and $1 \leq i \leq m$, where $S^{*} \subset S$ is a set of $m$ points, $m \leq n$, arranged in ascending order which configures a set of $m$ line segments that preserve the main characteristics of $S$. This set of $m$ line segments must describe the original shape as accurated as possible and using the minimum of information. This accuracy is given by the approximation error $\epsilon$ that exists between $S$ and $S^{*}$. Whereas the original set $S$ has an error equal to zero, once that in a discreet case, it owns all shape information, the resulting error from the polygonal approximation formed by $S^{*}$ can be calculated by the methodology used in \cite{journals/prl/KolesnikovF03,journals/pr/Yin03}. In this work, we compute the polygonal approximation error as the square sum of the perpendicular distance from each point to the line segment formed by $S^{*}$. In \cite{journals/pr/Yin03} the error is defined as

\begin{equation}
E(S,P)= \sum^{m}_{i=1} e(\widehat{x_{v(i)}x_{v(i+1)}}, \overline{x_{v(i)}x_{v(i+1)}}),
\end{equation}

where $v(m+1)=v(1)$ and $e(\widehat{x_{v(i)}x_{v(i+1)}}, \overline{x_{v(i)}x_{v(i+1)}})$ is the approximation error between arc $\widehat{x_{v(i)}x_{v(i+1)}}$ and the line segment $\overline{x_{v(i)}x_{v(i+1)}}$. $P$ denotes the set of $m$ line segments formed by the points in $S^{*}$:

\begin{equation}
P = \left\{ \overline{x_{v(1)}x_{v(2)}}, ,...,\overline{x_{v(m-1)}x_{v(m)}},\overline{x_{v(m)}x_{v(1)}} \right\}
\end{equation}

\begin{figure*}[!htb]
  \centering
  \includegraphics[width=0.75\textwidth]{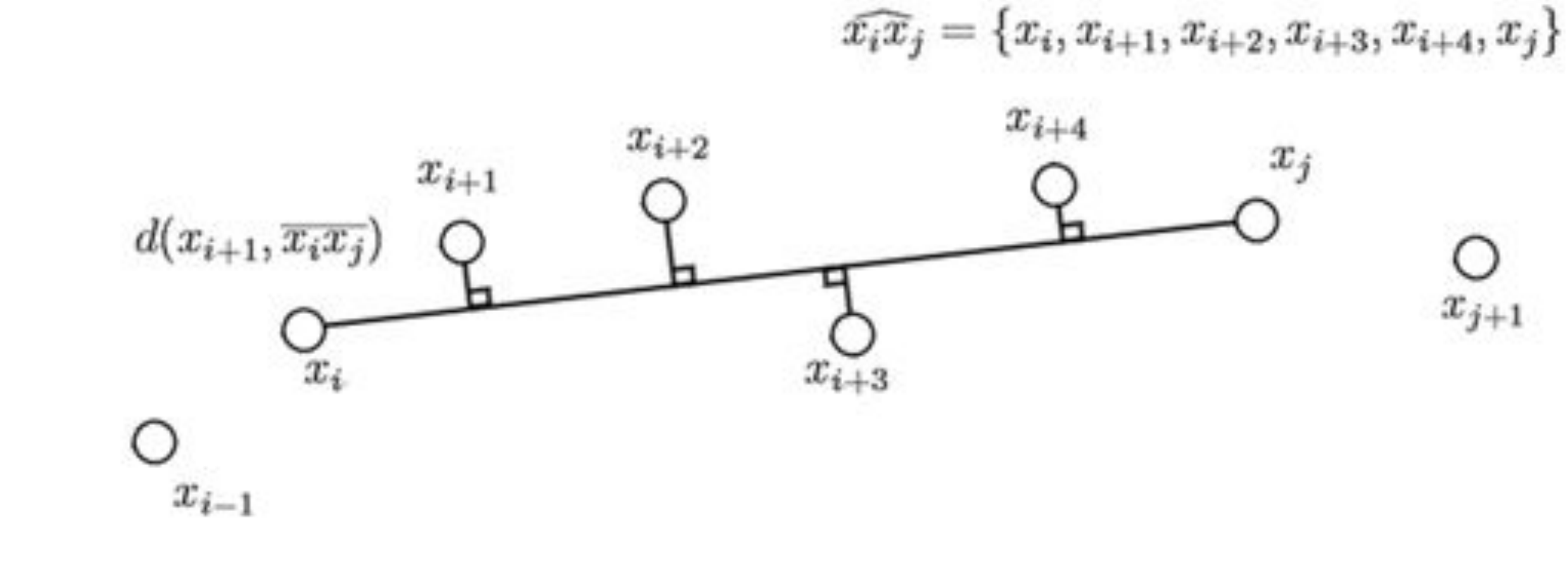}
  \caption{Example of arc $\widehat{x_{v(i)}x_{v(j)}}$ and computation of e$ (\widehat{x_{v(i)}x_{v(j)}}, \overline{x_{v(i)}x_{v(j)}})$. Adapted from \protect \cite{journals/pr/Yin03}.}
   
  \label{fig:calculo_arco}
\end{figure*}

The value of $e(\widehat{x_{v(i)}x_{v(i+1)}}, \overline{x_{v(i)}x_{v(i+1)}})$ is given by the equation defined by \cite{journals/pr/Yin03}, also used in the work of \cite{journals/prl/KolesnikovF03} and \cite{bb16649},

\begin{equation}
e(\widehat{x_{v(i)}x_{v(i+1)}}, \overline{x_{v(i)}x_{v(i+1)}}) = \sum^{}_{x_{j}\in \widehat{x_{v(i)}x_{v(i+1)}}} d^2(x_{j},\overline{x_{v(i)}x_{v(i+1)}}),
\end{equation}

where $d(x_{j},\overline{x_{v(i)}x_{v(i+1)}})$ is the perpendicular distance from point $x_j$ to its corresponding segment, $\overline{x_{v(i)}x_{v(i+1)}}$. Figure \ref{fig:calculo_arco} shows an arc $\widehat{x_{i}x_{j}}$ as a collection of six points, $\left\{ x_{i}, x_{i+1}, x_{i+2}, x_{i+3}, x_{i+4}, x_{j} \right\}$ and we compute the error of $\widehat{x_{v(i)}x_{v(i+1)}}$ as the sum of $d^{2}(x_{i},\overline{x_{v(i)}x_{v(i+1)}})+d^{2}(x_{i+1},\overline{x_{v(i)}x_{v(i+1)}})+d^{2}(x_{i+2},\overline{x_{v(i)}x_{v(i+1)}})+d^{2}(x_{i+3},\overline{x_{v(i)}x_{v(i+1)}})+d^{2}(x_{i+4},\overline{x_{v(i)}x_{v(i+1)}})+\\d^{2}(x_{j},\overline{x_{v(i)}x_{v(i+1)}})$.

\subsection{Complex Network model}
\label{sec:model}

In order to apply the Complex Network theory to the problem, a representation of the problem should be built like a graph $ G = \left\langle N, E\right\rangle $. We consider each point of the curve $S$ as a vertex in the graph, i.e., $ N = S$. This first approach is similar to Yin's Ant Colony Search method (ACS) for polygonal approximation \cite{journals/pr/Yin03}. The difference remains in the fact that ACS method computes the polygonal approximation as an optimization problem, while in this paper, properties from Complex Network theory are used in order to solve the polygonal approximation problem.

We compute the set of edges \textit{E} that composes the network in the following way: for each vertex $ x_ {i} \in S $, we examine the remaining vertices, $ x_ {j} \in S $, in clockwise order. We add a directed edge binding those vertices, $ \stackrel {\rightarrow} {x_ {i} x_ {j}} $, and the weight of that edge is the error of existing approximation between the arch $ \widehat {x_ {i} x_ {j}} $ and the segment of line $ \overline {x_ {i} x_ {j}} $, $e = (\widehat {x_ {i} x_ {j}}, \overline {x_ {i} x_ {j}}) $. 

The initial network is composed by a set of edges \textit{E} binding each pair of vertices of the network. This means that the network has regular behavior, once all vertices have the same number of connections. However, a regular network is not considered a Complex Network. It does not present any relevant property for the proposed application. So, it is necessary to transform this regular network into a Complex Network which owns relevant properties for the application.

Analyzing the Complex Network models existing in the literature, we noticed that the Small-World model proposed by Watts and Strongatz \cite{watts98} is the one which better fits into the polygonal approximation problem. This model is characterized by the existence of two basic properties in the network: (i) high clustering coefficient and (ii) small-world property. The clustering coefficient of a network is a measure of how interconnected its vertices are. It is defined as \begin{center} $ C = 3N_{\Delta} / N_{3}$, \end{center} where $ N_{\Delta}$ is the number of existing triangles in the network and $ N_{3} $ the number of connected triples. We consider a connected triple when a vertex $i$ is connected to a vertex $j$, and this vertex $j$ is connected to vertex $k$. A triangle occurs when vertices $i$ and $k$ are also connected. This is an interesting property, once the more the vertices are interconnected, the more paths between a pair of vertices are possible. The small-world property is a consequence of a high clustering coefficient. It is characterized by the existence of a small average geodesic path in the network. Literature describes a geodesic path as the shortest path connecting two vertices $i$ and $j$, where $i \neq j$. Therefore, these properties guarantee the existence of short paths among network vertices, and these short paths represent possible polygonal approximation of the curve.

The existence of these properties in the network is essential for attaining good results in polygonal approximation, and it is necessary to model the digital curve as a Small-World network. To construct a small-word network, we start with a regular lattice of $N$ vertices, where each vertex is connected to $\kappa$ vertices, and $N > \kappa > \log N > 1$. We randomly rearrange each edge with a probability $p$. When $p = 0$, we have an ordered lattice with high number of loops but large distances. When $p \rightarrow 1$, the network becomes a random graph with short distances but no loops. Watts and Strogatz have shown that, in an intermediate regime, both short distances and large number of loops are present. 

Therefore, for graph $G$, a regular lattice can be attained by selecting the $\kappa$ closest vertices of each vertex. Considering that the probability $p$ is straightly related to the approximation error between the arc $\widehat{x_{i} x_{j}} $ and the line segment $ \overline{x_ {i} x_ {j}}$, we built a Small-World network where the arcs with small error have prevailed. The probability $p$ of a vertex $i$ to be connected to a vertex $j$ is defined as:

\begin{equation}
p = 1 - \frac{e_{ij}}{\max(E)},
\end{equation}

where, $e_{ij}$ is the weight of the edge that connects the vertex $i$ to $j$. 
	
For a high value of $\kappa$ and a maximum error $\epsilon_{1}$, the rearrangement of edges occurs in a small interval and only whether $e_{ij} \leq \epsilon_{1}$. This edge rearrangement is similar to apply a threshold $\epsilon_{1}$ to the edge set $E$. 

In the light of this, the proposed idea of this work is: given a contour modeled as a regular network $G$, we transform $G$ into a Small-World network by applying a $\delta$ transformation over $E$, using different values of $\epsilon_{1}$. We apply this operation, henceforth represented as $\widehat{E} = \delta_{\epsilon_{1}}(E)$, to each edge $e_{ij}$, $e_{ij} \in E$, yielding a subset of edges $\widehat{E}$, $\widehat{E} \subseteq E$. So:

\begin{equation}
\delta_{\epsilon_{1}}(E) = \left\{ e_{ij}: e_{ij} \in E \wedge e_{ij} \leq \epsilon_{1} \right\}
\end{equation}

\begin{figure*}[!htb]
\centering
	\includegraphics[width=0.75\textwidth]{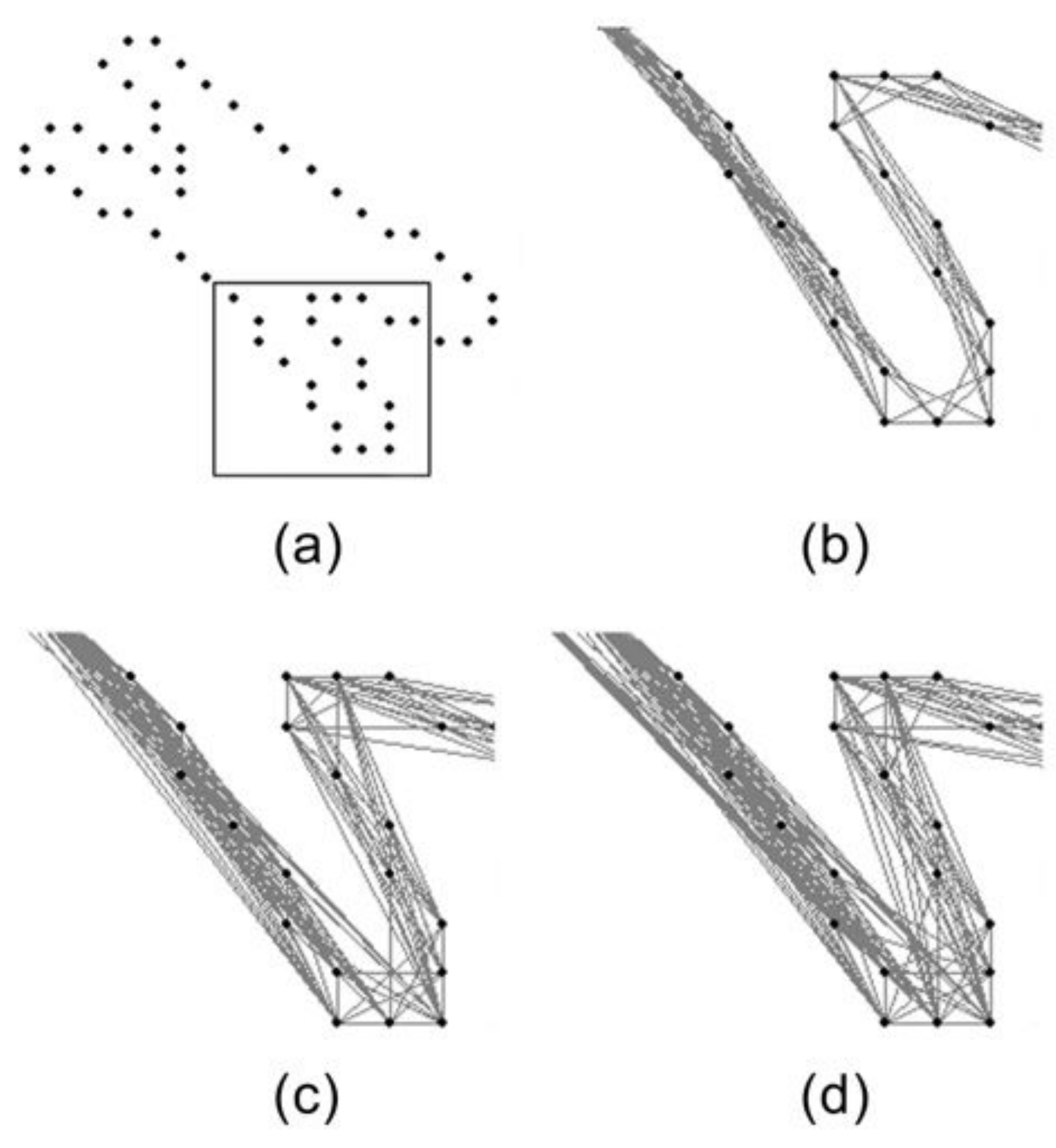}
	\caption{vertex visibility for several $\epsilon_{1}$: (a) Network vertices and zoom in area; (b) $\epsilon_{1} = 1$; (c) $\epsilon_{1} = 5$; (d) $\epsilon_{1} = 9$.}
	\label{fig:vertice_visibility}
\end{figure*}

We perform various $\delta$ transformations using different threshold $\epsilon_{1}$, which we increase in a regular interval. In this case, $\epsilon_{1}$ works as a visibility control of the network vertices, i.e., it limits the number of vertices that a vertex can reach to compose the approximation of an arc. Figure \ref{fig:vertice_visibility} shows how the vertex visibility changes for different values of $\epsilon_{1}$. As a consequence, it is possible to convert a regular network into a Complex Network that fits in the Small-World model proposed by Watts and Strongatz \cite{watts98}.

Once we model the contour as a Small-World network, a possible solution for the polygonal approximation problem is the set of vertices throughout a path formed by a set of edges \textit{E*}, which starts and ends in the same vertice, and that respects an error $\epsilon$ attributed to the polygonal approximation. Therefore, we restricted the polygonal approximation problem to find the geodesic path (between \textit{i} and \textit{i}) existing inside the network, respecting error $\epsilon$. For a better representation of the problem, some notations were defined. Consider $cycle_{v}$, the shortest cycle present in the network that begins and ends in vertex \textit{v}, the number of vertices in $cycle_{v}$ as $\left| cycle_{v} \right|$, and the error between the original curve \textit{S} and the corresponding approximation $cycle_{v}$ as $E(S,cycle_{v})$.

\subsection{Shortest cycle calculation}
\label{sec:menorciclo}

In a Complex Network, given two vertices \textit{s} and \textit{v} the geodesic path problem (or shortest path) consists of minimising the cost of crossing the graph between those two vertices. We define this cost as the sum of the weights of each edge traveled or just the number of traveled edges. The geodesic path between two vertices is an important rule of characterization of the internal structure of a network \cite{Boccaletti+2006}.

Specialized algorithms are well known in the literature for solving the shortest path problem. One of the most used methods is the Bellman-Ford algorithm, an iterative algorithm that uses the technique of edge relaxation \cite{cherkassky94shortest} to find minimum paths between an origin \textit{s} and a destiny \textit{v} in a network. For the yielded shortest path to be considered as a valid polygonal approximation, it is necessary to compute also the polygonal error of this. Therefore, we developed an adapted version of the Bellman-Ford algorithm which computes the shortest path in a network, minimizing also its polygonal error. In this work, we opted for using the Bellman-Ford algorithm due to its simplicity, easy implementation and generality.

Given a weighed and directed graph $G_{\epsilon_1} = \left\langle N,\widehat{E}\right\rangle$, an origin \textit{s} and a weight function \textit{w}, the algorithm returns the shortest path existing in the network, whose origin is the vertex \textit{s}, as well as the respective number of vertices and polygonal error of this way. For each vertex $v \in N$, the attribute \textit{d[v]} corresponds to the number of vertices in the shortest path between \textit{s} and \textit{v}. The attribute \textit{r[v]} specifies the vertex predecessor to vertex \textit{v}. We also introduced an attribute \textit{er[v]}, which specifies the error in the polygonal approximation computed from \textit{s} to \textit{v}. The process of initialization of the minimum path occurs according to the following algorithm:

\begin{tabular}{l}
\textit{procedure Initialize($G_{\epsilon_1}$,s)}\\
\textit{For each }$v \in N$\\
\quad \quad \textit{d[v]=}$\infty$\\
\quad \quad \textit{r[v]=NIL}\\
\quad \quad \textit{er[v]=$\infty$}\\
\textit{d[s]=0}\\
\textit{er[s]=0}
\end{tabular}

In the process of relaxation of edge \textit{(u, v)} we check whether the minimum path of a vertex \textit{v} and its respective error can be reduced by using a predecessor vertex \textit{u}. This stage consists of trying to decrement the error and the cost of the path estimated for \textit{v}, \textit{er[v]} and \textit{d[v]} respectively, and update its predecessor, \textit{r[v]}. This stage is done through the algorithm:

\begin{tabular}{l}
\textit{procedure relax(u,v,w)}\\
\textit{IF d[v] $>$ d[u] + 1}\\
\quad \quad \textit{d[v] = d[u] + 1}\\
\quad \quad \textit{r[v] = u}\\
\quad \quad \textit{er[v] = er[u] + w(u, v)}\\
\textit{ELSE}\\
\quad \quad \textit{IF (d[v] $=$ d[u] + 1) AND (er[v] $>$ er[u] + w(u, v))}\\
\quad \quad \quad \quad \textit{er[v] = er[u] + w(u, v)}\\
\quad \quad \quad \quad \textit{r[v] = u}\\
\end{tabular}

Finally, we describe the adapted algorithm of Bellman-Ford as the following set of steps:

\begin{tabular}{l}
\textit{BELLMAN-FORD($G_{\epsilon_1}$,s,w)}\\
\textit{Initialize($G_{\epsilon_1}$,s)}\\
\textit{For each }$v \in N$\\
\quad \textit{For each } $(u,v) \in \widehat{E}$\\
\quad \quad \textit{Relax(u,v,w)}\\
\end{tabular}

In the polygonal approximation case, it is necessary to achieve the geodesic path in a Complex Network that starts in a vertex \textit{s} and ends in a vertex \textit{v}, where $v = s$. We consider this geodesic path as the shortest cycle of the network starting in vertex \textit{s}.

\subsection{Hybrid strategy}
\label{sec:hybrid strategy}

An algorithm of shortest path calculation often lead the local minimums before reaching the global optima. Some researchers suggest that using a hybrid strategy between the iterations of the global search method as a way to minimize the possibility of convergence of the search into local minimums \cite{Goldberg1989}. This assures the possibility of obtaining excellent results. In the light of this, we decided to use a heuristic of local search in SP (referred to as Hybrid SP or HSP).

In polygonal approximation done for \cite{peng2004}, heuristics of split-and-merge are adopted. The method is based on the use of the hierarchical segment-merging method \cite{Leu1988}, where it is assumed that adjacent segments $\widehat{x_{v(i)}x_{v(i+1)}}$ and $\widehat{x_{v(i+1)}x_{v(i+2)}}$ have been tested and if the distance between the furthest point of $\widehat{x_{v(i)}x_{v(i+2)}}$ and the line segment $\overline{x_{v(i)}x_{v(i+2)}}$ is smaller than error tolerance, the two segments can be merged yielding the new segment $\widehat{x_{v(i)}x_{v(i+2)}}$. The algorithm proceeds analogously by observing whether the newly defined segment $\widehat{x_{v(i)}x_{v(i+2)}}$ and the subsequent segment $\widehat{x_{v(i+2)}x_{v(j)}}$ can be merged. These steps proceed iteratively until the last two segments have been tested. Then, the last and first segments can finally be tested. In addition, \cite{peng2004} adopts another local optimizer called the segment-adjusting-and-merging technique where it examines each vertex of the polygon in turn and, if possible, it is adjusted to a new data point on the arc between the two adjacent vertices, so that the approximation error is reduced as much as possible. The execution stops when the dominant point stack has not changed (i.e. new dominant points and segments have not been merged).

\subsection{Algorithm}
\label{sec:algorithm}

In order to get a polygonal approximation which better characterizes a curve, we proposed the algorithm described in Table \ref{tab:algoritmo}. The basic idea is to select a set of edges $\widehat{E}$ from the network, where the weight of each edge is not larger than a threshold $\epsilon_{1}$. This new set of edges transforms a regular network in a Small-World Complex Network, changing its behavior and properties. From this new network, the shortest cycle calculation takes place using the algorithm described in Section \ref{sec:menorciclo}. Once the initial vertex can carry out some influence on the calculated cycle, we consider each network vertex as a possible starting vertex for the cycle.

\begin{center}
\begin{table*}[!ht]
\centering
\begin{tabular}{|l|}
\hline
Input\\
$S = \left\{ x_{1},x_{2},...,x_{n}\right\}$: set of points of curve in clockwise order.\\
$\epsilon_{max}$: maximum value that can be attributed to $\epsilon_{1}$. \\
1. Initialize\\
\quad Build the directed graph  $G = \left\langle N,E\right\rangle$, as described in \ref{sec:model}.\\
\quad Be $shortest_{cycle} = \left\{ \overline{x_{1}x_{2}}, \overline{x_{2}x_{3}},..., \overline{x_{n-1}x_{n}},\overline{x_{n}x_{1}} \right\}$.\\
\quad Set $\epsilon_1 = 0$. \\
2. While $\epsilon_{1} \leq \epsilon_{max}$ \\
\quad $\epsilon_1 = \epsilon_1 + 0.1$. \\
\quad Select $\widehat{E}$.\\
\quad For each $v \in N$\\
\quad \quad Calculates the shortest cycle beginning in  \textit{v}, $cycle_{v}$.\\
\quad Be $current_{shortest\_cycle}$ the shortest cycle calculated with error smaller than $\epsilon$. \\
\quad If $\left| current_{shortest\_cycle} \right| < \left| shortest_{cycle} \right|$ \\ 
\quad \quad $shortest_{cycle} = current_{shortest\_cycle}$.\\
3. Improve the solution quality in $shortest_{cycle}$ using the local optimizer.\\
4. Display $shortest_{cycle}$.\\
\hline
\end{tabular}
\caption{Proposed procedure}
\label{tab:algoritmo}
\end{table*}
\end{center}

\section{Experiments and discussion}
\label{sec:results}

\begin{figure*}[!htbp]
\centering
	\includegraphics[width=0.65\textwidth]{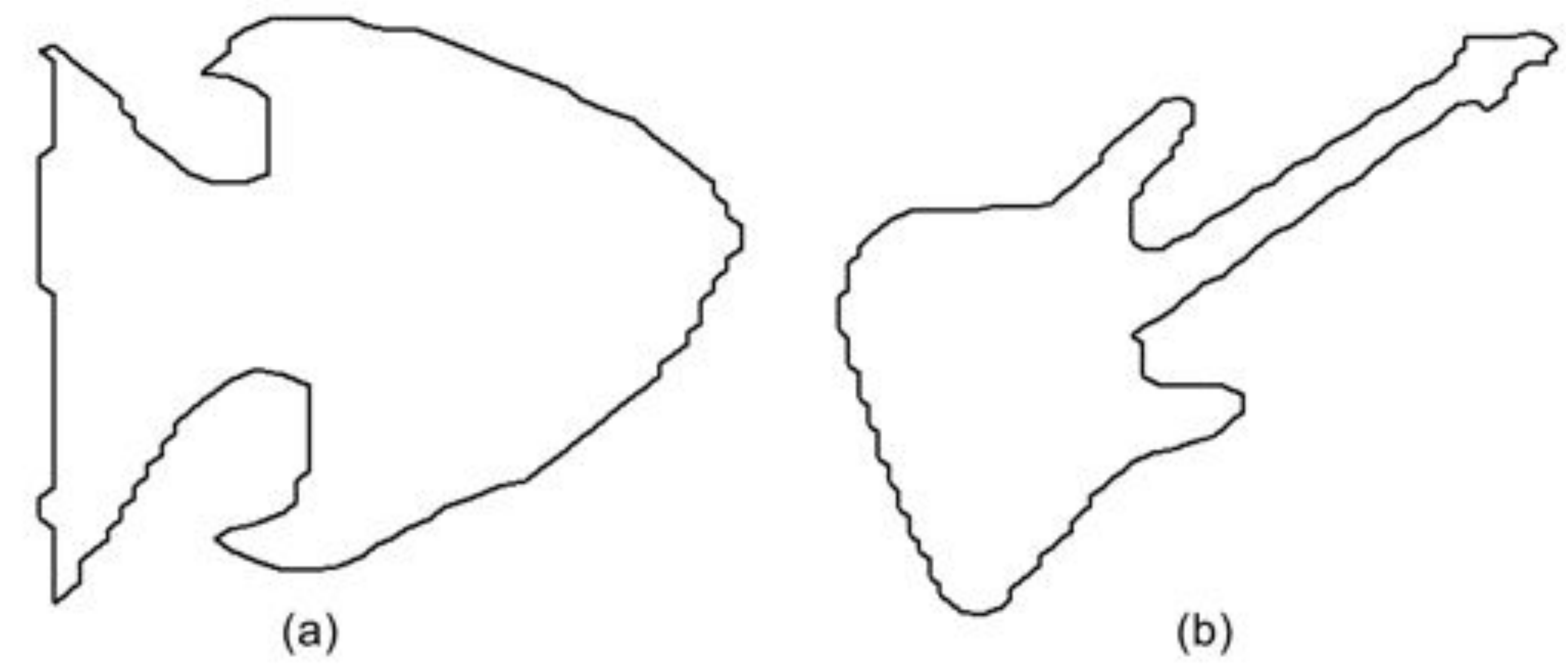}
	\caption{Real curves used to validate the proposed method: (a)Fish ($n = 185$) and (b)Guitar ($n = 193$).}
	\label{fig:curvas_usadas}
\end{figure*}

\begin{figure*}[!htbp]
\centering
	\includegraphics[width=0.75\textwidth]{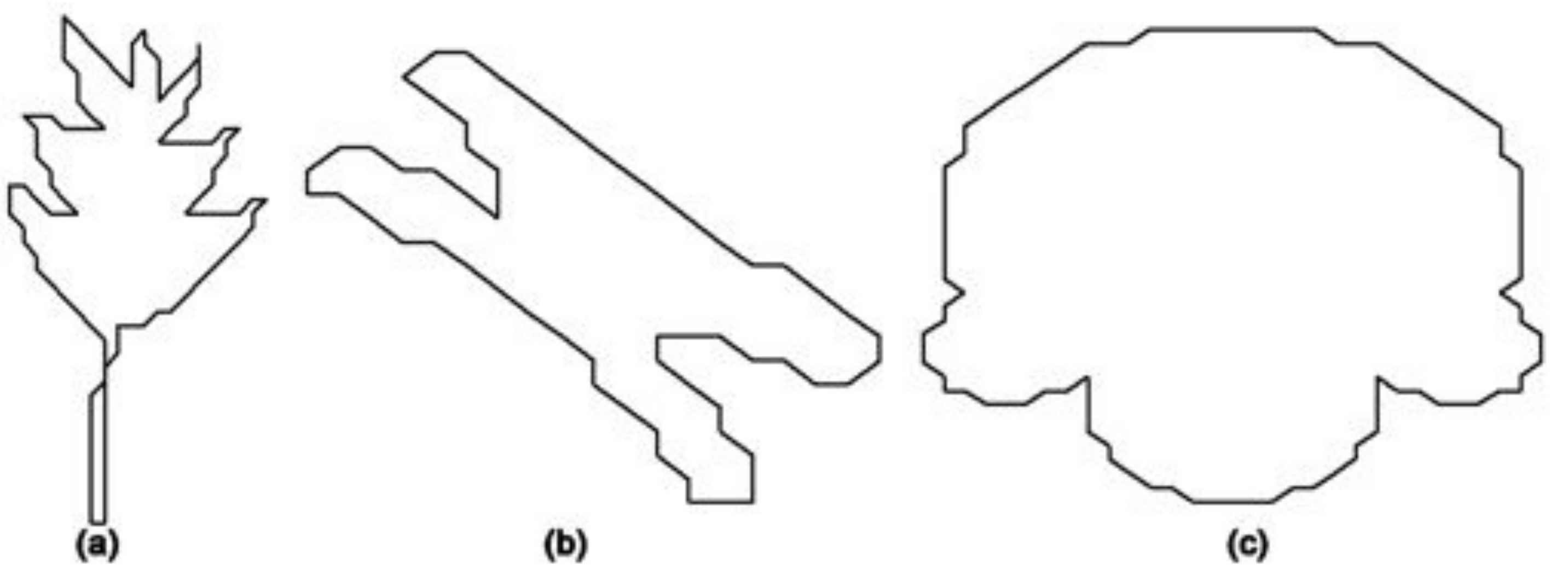}
	\caption{Benchmark curves used to validate the proposed method: (a)Leaf ($n = 120$), (b)Chromosome ($n = 60$) and (c) Semicircle ($n = 102$).}
	\label{fig:curvas_usadas2}
\end{figure*}

In order to observe the efficiency of the proposed method for curve polygonal approximation, we performed an experiment. For this experiment, we used two curves achieved from shape contour (Figure \ref{fig:curvas_usadas}). We also used three synthesized benchmark curves (Figure \ref{fig:curvas_usadas2}), which are broadly used in the literature \cite{held-1994,ray1995,teh-1989,peng2004}. These curves had their respective polygonal approximation calculated using the proposed method with different values for the parameter $\epsilon_{1}$. This was performed in order to observe the influence of $\epsilon_{1}$ on the approximation results, as well as the characteristics and particularities this proposed method presents.

\subsection{Parameters evaluation}

\begin{figure*}[!htbp]
\centering
	\includegraphics[width=0.75\textwidth]{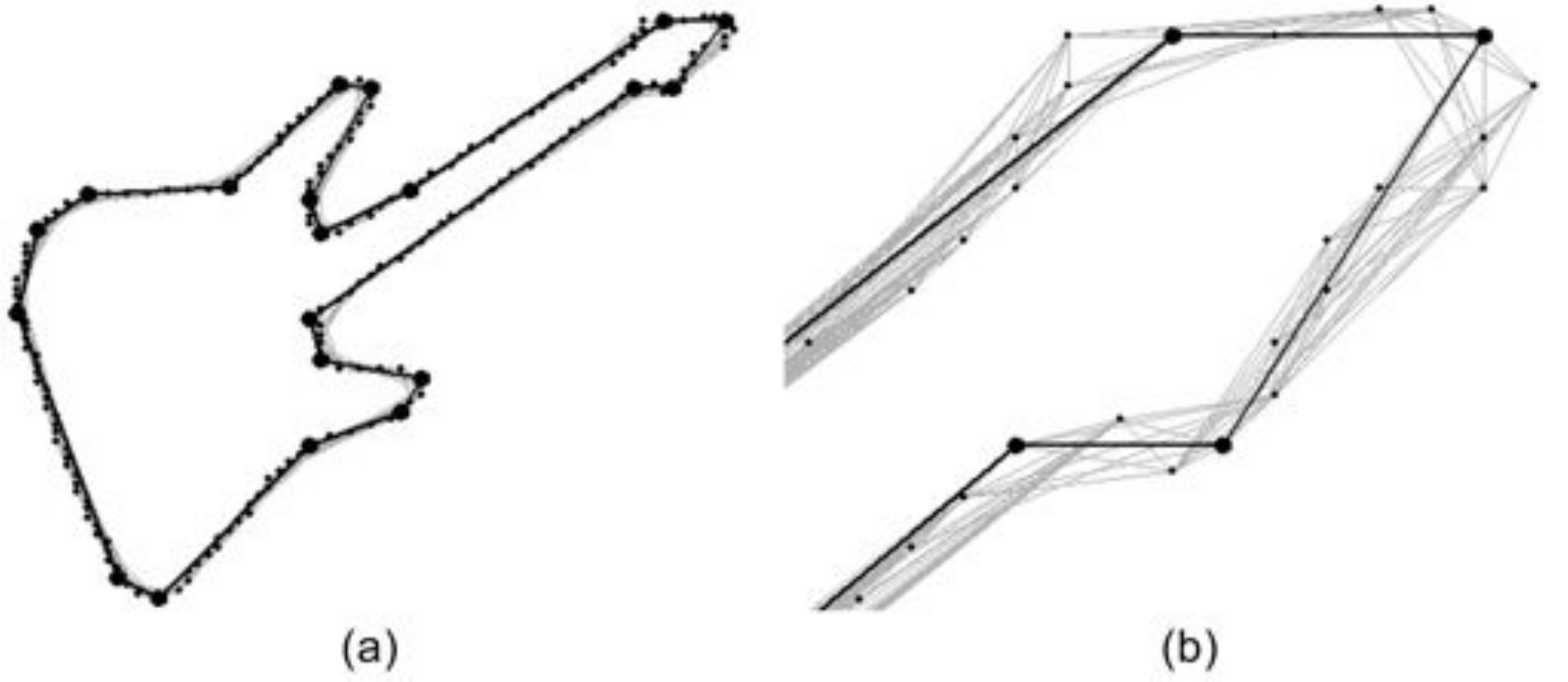}
	\caption{(a) Guitar figure modeled as a Complex Network, using $\epsilon_{1}=9$ and its shortest cycle. (b) Zoom in on a small part of the network.}
	\label{fig:redefolha}
\end{figure*}

\begin{figure}[!htbp]
\centering
	\includegraphics[width=0.5\textwidth]{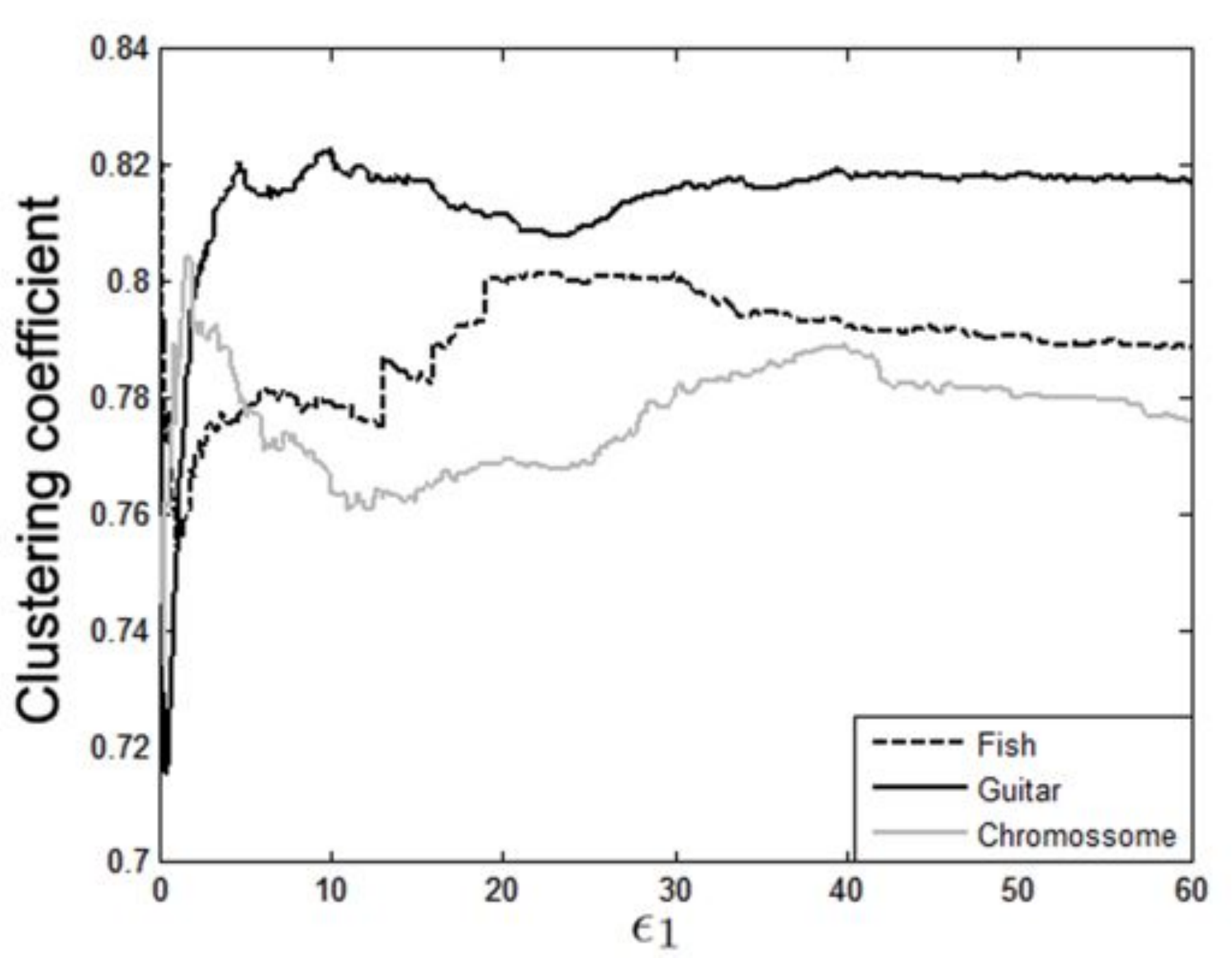}
	\caption{Clustering coefficient of the Fish Figure against $\epsilon_{1}$ value.}
	\label{fig:coefclustering}
\end{figure}

\begin{figure}[!htbp]
\centering
	\includegraphics[width=0.5\textwidth]{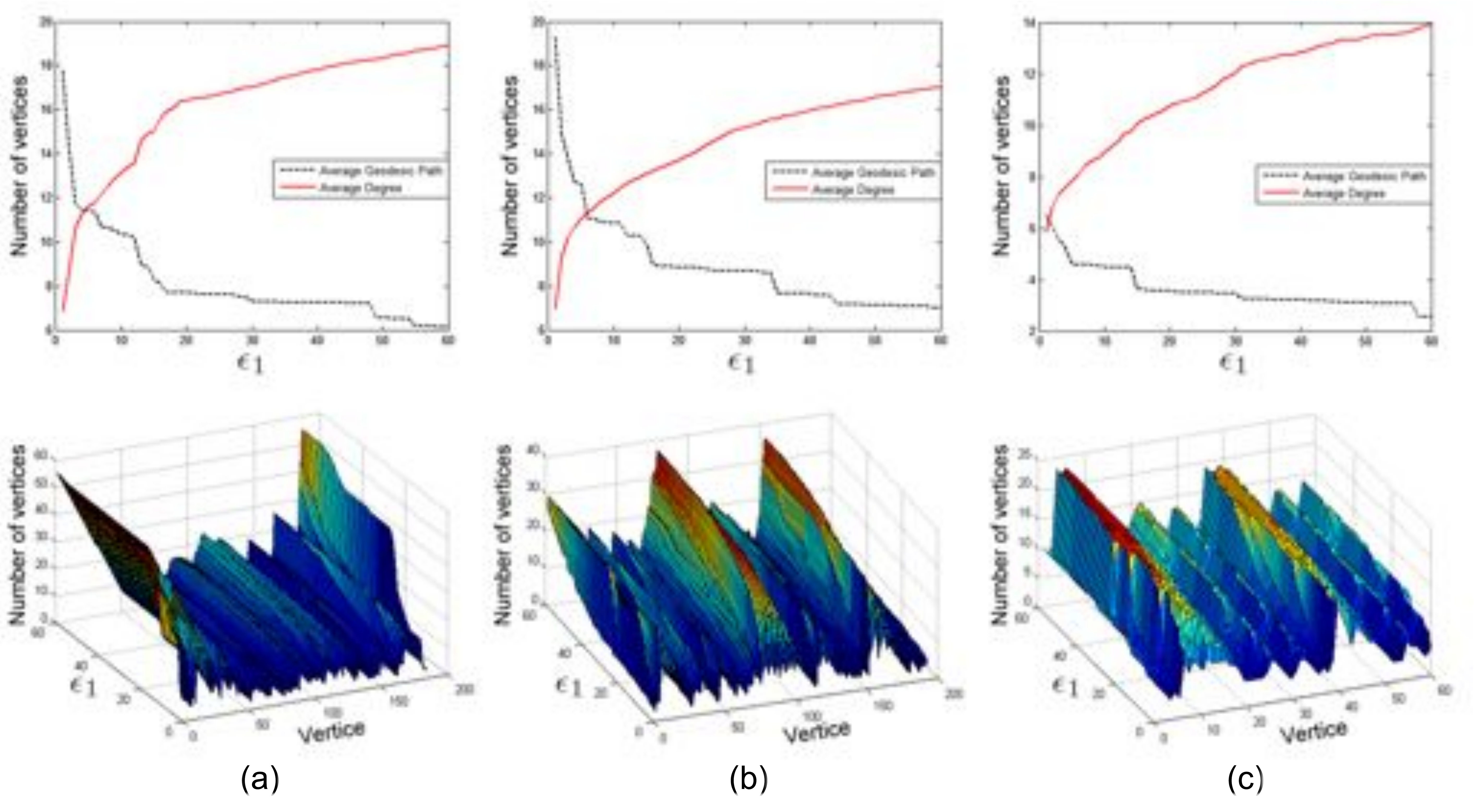}
	\caption{Variation of shortest cycle and average geodesic path length against $\epsilon_{1}$ value for the Fish Figure.}
	\label{fig:limiarerro}
\end{figure}

As previously described, parameter $\epsilon_{1}$ is able to convert the initial network (which presents a regular behavior) into a Small-World Complex Network (Figure \ref{fig:redefolha}). However, we need to observe if this network really owns the same properties that characterize a Small-World network. Figure \ref{fig:coefclustering} shows that the modeled network presents a high clustering coefficient, regardless of the value of parameter $\epsilon_{1}$. This shows a high level of transitivity between the network vertices, i.e., it is possible to find more than a path from vertex \textit{i} to vertex \textit{j}. The other property of Small-World networks refers to the existence of a small average geodesic path, i.e., every vertex can be attained from any other vertex through a small number of edges. This is an important property to the polygonal approximation problem, as the polygonal approximation is a geodesic path that starts and ends at the same vertice. Figure \ref{fig:limiarerro} shows the length variation of the average geodesic path and shortest cycle of the network against the value of parameter $\epsilon_{1}$. We notice that the shortest cycle length is roughly the double of the length of the average geodesic path. In fact, a cycle can be considered as the concatenation of two average geodesic paths: the path between vertices \textit{i} and \textit{k}, and the path between vertices \textit{k} and \textit{j}, where $i = j$.

\begin{figure*}[!htbp]
\centering
	\includegraphics[width=0.75\textwidth]{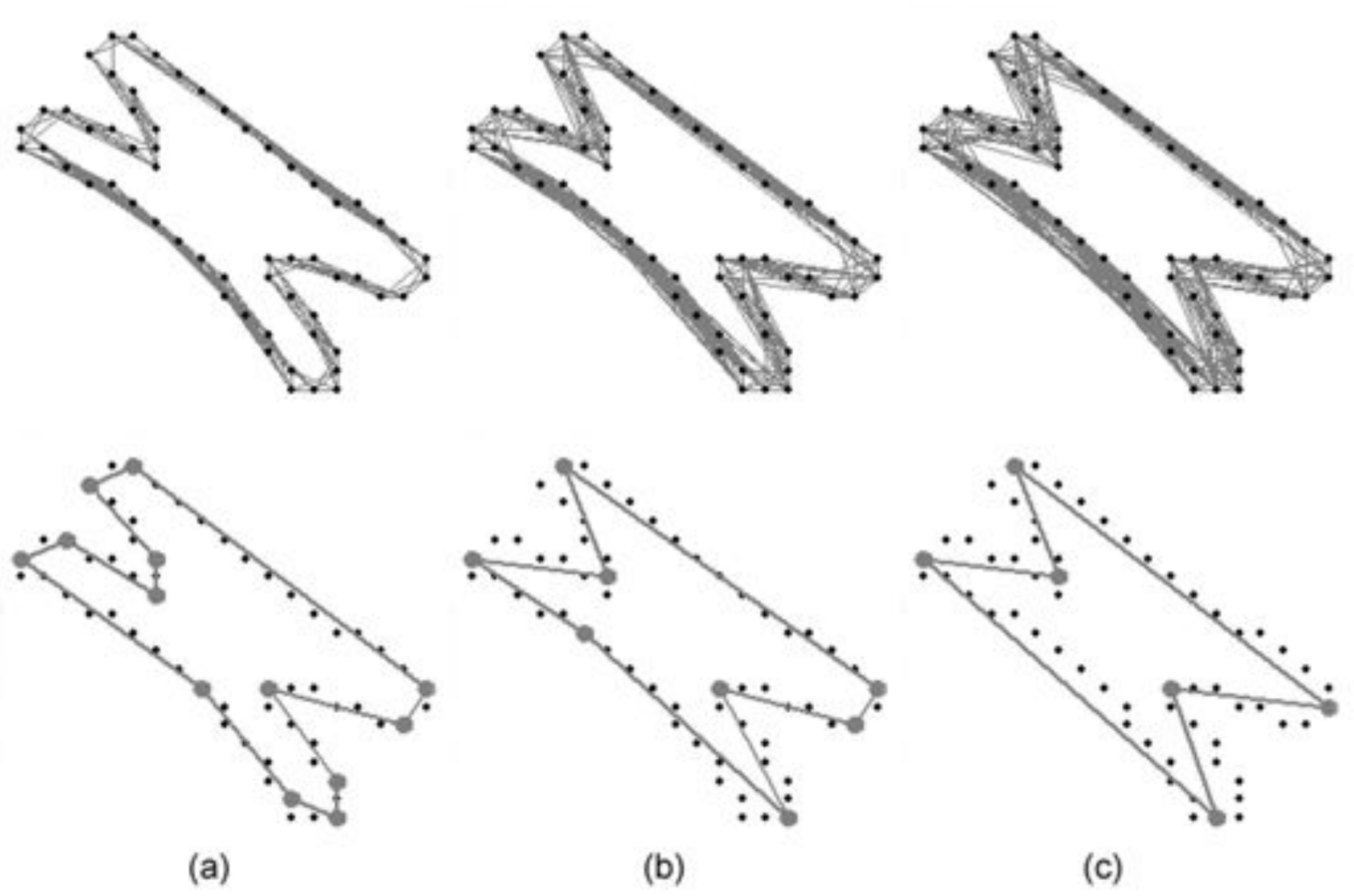}
	\caption{Network transitivity and best polygonal approximation for several $\epsilon_{1}$: (a) $\epsilon_{1} = 1$; (b) $\epsilon_{1} = 5$; (c) $\epsilon_{1} = 9$.}
	\label{fig:rede_threshold}
\end{figure*}

\begin{figure}[!htbp]
\centering
	\includegraphics[width=0.45\textwidth]{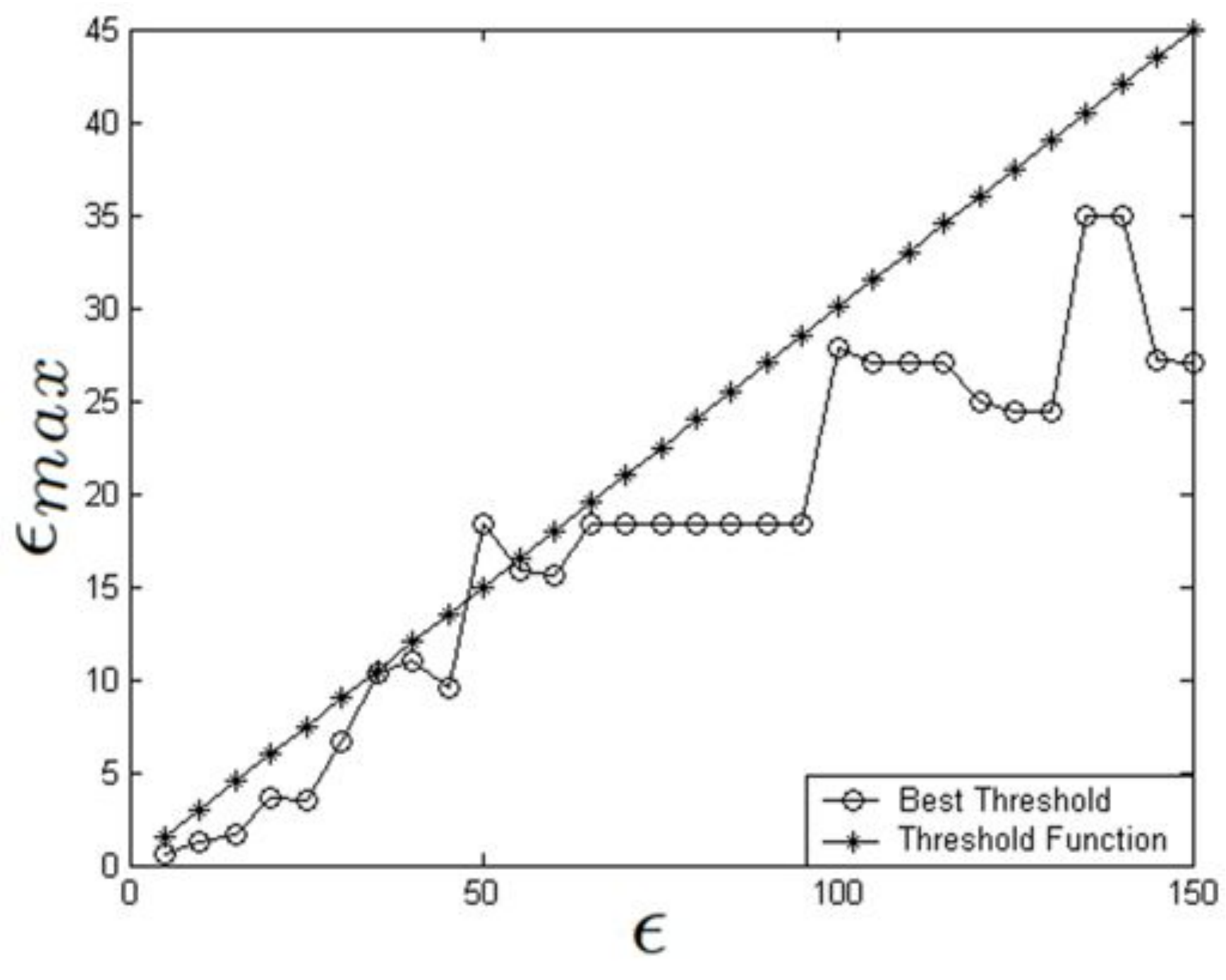}
	\caption{Maximum threshold ($\epsilon_{max}$) used for a specific $\epsilon$ and a possible function for it.}
	\label{fig:emax_function}
\end{figure}

An important aspect to be considered is the \textit{stop criterion}. As we increase the threshold, the network transitivity changes, as well as the maximum error allowed for a line segment $\overline{x_{i}x_{j}}$. This makes possible to find out new paths in the network, and most of them using a smaller number of vertices (Figure \ref{fig:rede_threshold}). In this sense, the $\epsilon$ parameter acts as a limit of the maximum error that a polygonal approximation can achieve. It prevents us to build polygonal approximations with an inadequate number of vertices and, as a result, it avoids the deformation of the original curve. 

From a specific value of $\epsilon_{1}$, the solution found by the method does not change, i.e., it remains the same independent of how much the value of $\epsilon_{1}$ increases (Figure \ref{fig:limiarerro}). We define this limit value for $\epsilon_{1}$, $\epsilon_{max}$, as the maximum threshold that can be applied to the network in order to achieve a polygonal approximation which respects the maximum error $\epsilon$.

When we analyze the network behavior for different $\epsilon$ values, we realize that the value of $\epsilon_{max}$ depends on the chosen $\epsilon$. In general, the value of $\epsilon_{max}$ can be assumed, roughly, as being $0.3 * \epsilon$. In some cases, like $\epsilon < 25$, $\epsilon_{max}$ value can be decreased to $0.2 * \epsilon$. On the other hand, $\epsilon_{max}$ is also dependent of curve aspect, and in some few case, a $\epsilon_{max} > 0.3 * \epsilon$ is necessary (Figure \ref{fig:emax_function}).

As the value of $\epsilon_{1}$ changes, other characteristics of the network also change. An interesting characteristic is the degree presented for each vertex in the network. The degree is the number of connections that a vertex owns in the network, i.e., the number of neighbor vertices connected to it. It is a characteristic that exists in all networks, regardless of its model, and it is related to network transitivity. Figure \ref{fig:rede_grau} shows how the vertex degree changes as we increase $\epsilon_{1}$. This figure also shows the vertices selected to compose a polygonal approximation. We notice that most of the selected vertices remain close to local maximum points of the curve, i.e., these vertices exist in a high transitivity region of the network. They correspond to the points, in the original contour, where the higher changes in curve orientation occur. The connection that exists between the original contour and the modelled network is clear, where the curvature point in the original contour is directly related to the vertex degree.

\begin{figure*}
\centering
	\includegraphics[width=0.95\textwidth]{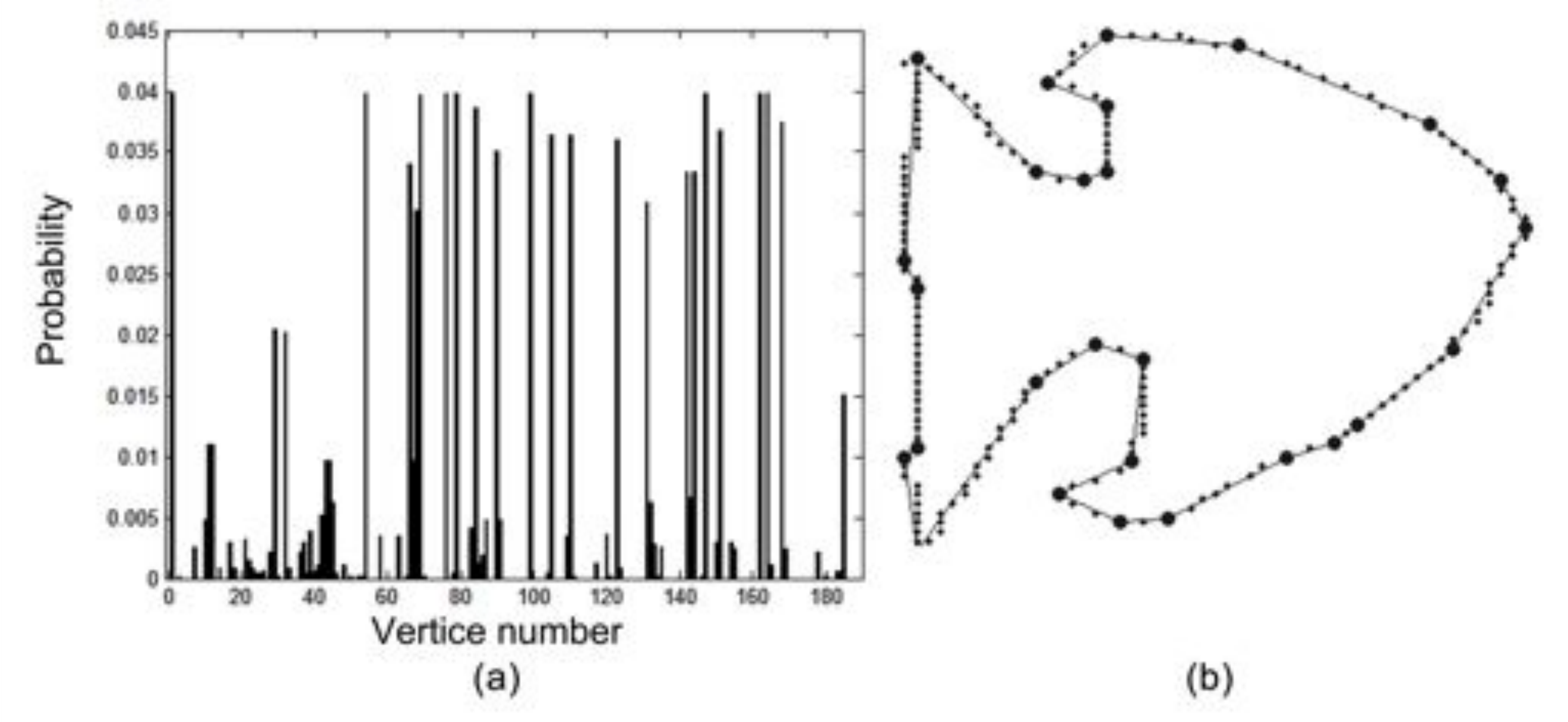}
	\caption{(a)Probability of a vertex to be chosen to compose a shortest cycle of the Fish Figure, using $\epsilon_{1}=2.5$. (b) Approximation composed by the vertex with higher a probability.}
	\label{fig:probvertices}
\end{figure*}

An important detail about the method is the initial vertex used in the shortest cycle calculation. By analysing the shortest paths achieved when we consider each network vertex as a possible initial vertice, we realize that the initial vertex has no influence on the approximation result. Indeed, regardless of the initial vertex, the network presents some vertices with a higher probability to be chosen to compose the shortest cycle (Figure \ref{fig:probvertices}). This indicates that there is a set of vertices in the network where the cost to cross the network is minimum. Parameter $\epsilon_{1}$ is responsible for the existence of this set of vertices, as it limits the visibility of the network vertices and it prevents the construction of line segments with a high approximation error. Another consequence of this network behaviour is the small variation of the calculated cycle, regardless of the initial vertex. 

\begin{center}
\begin{figure*}[!ht]
\centering
	\includegraphics[width=0.95\textwidth]{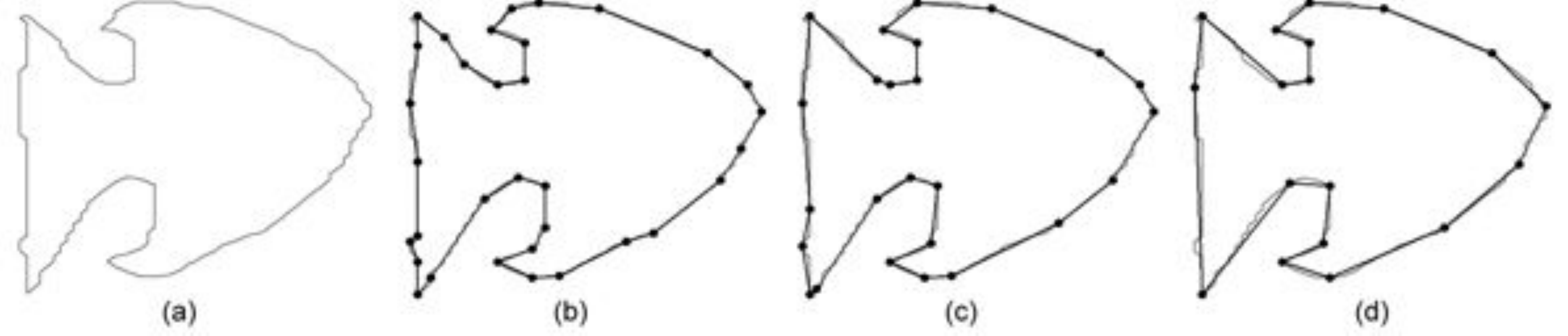}
	\caption{Examples of polygonal approximation for the Fish Figure, with respective approximation error ($\epsilon$) and number of vertices (\textit{m}): (a) Original Curve; (b) $\epsilon = 15.24$, $ m = 33$; (c) $\epsilon = 24.84$, $ m = 25$; (d) $\epsilon = 56,54$, $ m = 18$}
	\label{fig:resultadosfolha}
\end{figure*}
\end{center}

\begin{figure*}[!ht]
\centering
	\includegraphics[width=0.95\textwidth]{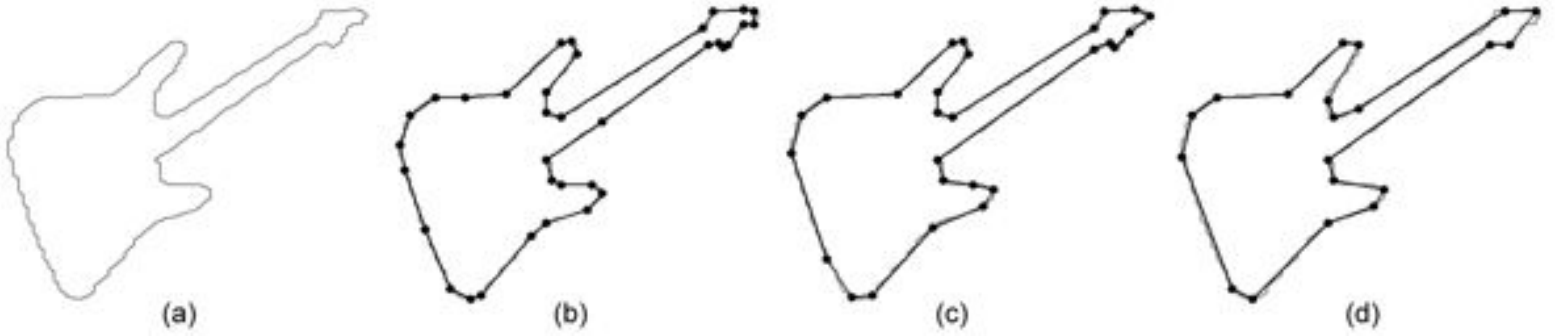}
	\caption{Examples of polygonal approximation for the Guitar Figure, with respective approximation error ($\epsilon$) and number of vertices(\textit{m}): (a) Original Curve; (b) $\epsilon = 18.72$, $ m = 35$; (c) $\epsilon = 30,79$, $ m = 27$; (d) $\epsilon = 56.90$, $ m = 20$ }
	\label{fig:resultadosplane}
\end{figure*}

Figures \ref{fig:resultadosfolha} and \ref{fig:resultadosplane} present some results for polygonal approximation according to the value of parameter $\epsilon_{1}$. As the value of $\epsilon_{1}$ increases, the number of vertices used to compose the polygonal approximation decreases and, consequently, the error for this polygonal approximation increases. 

\subsection{Comparison with other methods}

In order to improve the evaluation of our method, we compared our approach with other polygonal approximation method from literature. For this experiment we used many sophisticated polygonal approximation methods: the Discrete Particle Swarm Optimization (DPSO) \cite{peng2004}, the Yin's Ant Colony Search method (ACS) \cite{journals/pr/Yin03}, the Dominant Point Deletion (DPD) \cite{bb21110}, the Break Point Suppression (BPS) \cite{bb21121}, vertex Betweeness \cite{journals/isci/BackesB13} and using adaptive optimizations of the sub-set of dominant points of the curve (AO) \cite{journals/prl/ParvezM10}. It is important to emphasize that we implemented and run the methods used in this experiment according to the proposed specifications in their respective papers. Taking this into account, we can make a fair analysis and we avoid making a bias of the results of a specific method.

Table \ref{tab:resRCPS} shows the comparative performances among the various methods used in this experiment. For this experiment, we used our Shortest Path Complex Network method (SP) and a hybrid version (HSP). For the DPSO method \cite{Kennedy_Eberhart97}, we also compared its hybrid versions (referred to as HDPSO1 and HDPSO2) \cite{peng2004}. We evaluated each method by using different maximum errors $\epsilon_{max}$ per curve in order to test its efficiency to build polygons. We also conducted the methods to minimize the number of polygonal points, $m$.

For some curves and error tolerances, the Hybrid SP version is superior to the SP method. The hybrid strategy helps the SP method to explore regions which were despised in the execution. As we increase the threshold value $\epsilon_{1}$ (which is related to vertice visibility), the SP method privileges vertices which are more distant to compose the polygonal approximation, despising nearby vertices. The optimization used on HSP re-evaluate some despised vertices and it improves the results achieved by the method. We notice that the SP and HSP methods overcome other methods. However, evaluating an method only based on $\epsilon$ and $m$ does not provide enough information about its performance. Rosin \cite{bb16661} proposes to evaluate the performance of a method using an analysis of both characteristics. This measure is called \textit{Merit} and it is defined as:
\begin{equation}
Merit = 100 \times \sqrt{Fidelity \times Efficiency},
\end{equation}
where
\begin{equation}
Fidelity = \frac{\epsilon_{opt}}{\epsilon_{approx}}
\end{equation}
and
\begin{equation}
Efficiency = \frac{m_{opt}}{m_{approx}}.
\end{equation}

The merit is based on the fidelity and efficiency measures, which are computed from the results of the evaluated and an optimal algorithms. Fidelity measures how well the approximating polygon fits the curve in terms of the approximation error, while Efficiency measures how compact the approximating polygon is. Parameters $\epsilon_{opt}$ and $\epsilon_{approx}$ are, respectively, the approximation error produced by optimal and tested algorithms when we produce a polygon with the same number of vertices, while $m_{opt}$ and $m_{approx}$ are the number of vertices from the polygon produced by, respectively, optimal and evaluated algorithms for the same approximation error. Since it is difficult to compute a polygon for a specific approximation error, $m_{opt}$ is estimated by linear interpolation of the two closest values of $m$. We computed the optimal polygons using the dynamic programming approach \cite{journals/pr/Sato92}. Table \ref{tab:resRCPS} shows the merit of the approximating polygons produced by various algorithms. According to the merit analysis, both SP and HSP algorithms present higher values of merits according to our previous analysis.

\begin{table*}[!htbp]
	\centering
	\scriptsize 
	\setlength{\tabcolsep}{2.0pt}
		\begin{tabular}{|c|c|cc|cc|cc|cc|cc|}
		\hline
& & $\epsilon_{max} = 150$ & & $\epsilon_{max} = 100$ & & $\epsilon_{max} = 90$ & & $\epsilon_{max} = 30$ & & $\epsilon_{max} = 15$ & \\ \cline{3-12}
& Method & $m(\epsilon)$ & Merit & $m(\epsilon)$ & Merit & $m(\epsilon)$ & Merit & $m(\epsilon)$ & Merit & $m(\epsilon)$ & Merit\\
\hline
\multirow{8}{*}{\rotatebox{90}{\mbox{Leaf ($n = 120$)}}}
& DPSO   & 11.9(147.7) & 58.9 & 14.6(98.9) & 51.0 & 15.2(89.5) & 50.4 & 20.1(29.4) & 54.5 & 24.5(14.7) & 60.6\\
& HDPSO1 & 10.0(126.5) & 75.0 & 12.5(97.1) & 74.2 & 13.0(88.0) & 70.4 & 16.7(26.4) & 82.0 & 20.0(15.0) & 87.1 \\
& HDPSO2 & 10.0(126.5) & 75.0 & 12.0(93.3) & 83.5 & 12.5(82.6) & 80.4 & 16.0(26.6) & 92.9 & 20.0(15.0) & 87.1 \\
& ACS    & 11.2(149.5) & 55.6 & 13.0(88.6) & 81.1 & 13.2(81.8) & 65.2 & 17.2(21.3) & 86.2 & 22.2(15.0) & 81.4 \\
& BPS    & 16.0(44.5) & 71.7 & 16.0(44.5) & 71.7 & 16.0(44.5) & 71.7 & 20.0(15.4) & 90.2 & 23.0(14.2) & 75.1 \\
& DPD    & 10.0(127.2) & 87.0 & 12.0(85.9) & 80.3 & 12.0(85.9) & 80.3 & 16.0(27.0) & 98.8 & 20.0(14.1) & 96.1 \\
& AO		 & 18.0(33.1) & 64.2 & 18.0(33.1) & 64.2 & 18.0(33.1) & 64.2 & 19.0(23.9) & 74.0 & 21.0(14.5) & 87.9\\
& Betweenness&9.0(136.2) & 98.4 & 11.0(99.7) & 91.4 & 12.0(70.5) & 90.8 & 16.0(27.5) & 97.8 & 20.0(14.1) & 96.4 \\
& SP     & 9.0(147.2)& 82.1 & 11.0(100.0)& 96.8 & 12.0(65.38)& 93.1 & 16.0(27.09)& 98.9 & 20.0(14.84)& 95.3\\
& HSP    & 9.0(147.2)& 82.1 & 11.0(100.0)& 96.8 & 12.0(65.38)& 93.1 & 16.0(27.09)& 98.9 & 20.0(14.84)& 95.3\\
\hline
& & $\epsilon_{max} = 30$ & & $\epsilon_{max} = 20$ & & $\epsilon_{max} = 10$ & & $\epsilon_{max} = 8$ & & $\epsilon_{max} = 6$ & \\ \cline{3-12}
& Method & $m(\epsilon)$ & Merit & $m(\epsilon)$ & Merit & $m(\epsilon)$ & Merit & $m(\epsilon)$ & Merit & $m(\epsilon)$ & Merit\\
\hline
\multirow{8}{*}{\rotatebox{90}{\mbox{Chromosome ($n = 60$)}}}
& DPSO   & 6.7(29.4) & 75.1 & 8.6(17.5) & 65.3 & 11.6(9.5) & 65.0 & 12.4(7.7) & 71.7 & 14.4(5.9) & 72.7 \\
& HDPSO1 & 6.0(29.7) & 87.1 & 7.0(17.7) & 95.0 & 10.0(9.8) & 81.6 & 11.0(7.8) & 85.0 & 12.7(5.8) & 87.2 \\
& HDPSO2 & 6.0(29.7) & 87.1 & 7.0(17.7) & 95.0 & 10.0(9.8) & 81.6 & 11.0(7.8) & 85.0 & 12.0(5.8) & 92.9 \\
& ACS    & 6.0(27.8) & 70.4 & 8.0(19.8) & 87.3 & 10.0(9.5) & 87.2 & 11.0(7.8) & 91.2 & 12.8(5.7) & 87.5 \\
& BPS    & 9.0(18.8) & 69.7 & 9.0(18.8) & 69.7 & 13.0(8.5) & 65.6 & 15.0(6.1) & 69.2 & 17.0(3.8) & 85.7 \\
& DPD    & 6.0(23.7) & 100.0 & 7.0(17.7) & 100.0 & 10.0(8.0) & 100.0 & 11.0(7.6) & 93.8 & 12.0(5.8) & 100.0 \\
& AO		 & 7.0(29.8) & 65.1 & 8.0(19.7) & 75.4 & 11.0(7.8) & 92.1 & 11.0(7.8) & 92.1 & 12.0(5.8) & 100.0 \\
& Betweenness&6.0(23.7) & 100.0 & 7.0(19.9) & 91.7 & 10.0(8.1) & 99.9 & 11.0(7.8) & 92.0 & 12.0(5.8) & 100.0 \\
& SP     &6.0(25.98)& 88.1 & 7.0(19.76)& 99.7 & 10.0(8.79)& 98.5 & 11.0(7.46)& 100.0 & 12.0(5.97)& 98.8\\
& HSP    &6.0(25.98)& 88.1 & 7.0(19.76)& 99.7 & 10.0(8.79)& 98.5 & 11.0(7.46)& 100.0 & 12.0(5.97)& 98.8\\
\hline
& & $\epsilon_{max} = 60$ & & $\epsilon_{max} = 30$ & & $\epsilon_{max} = 25$ & & $\epsilon_{max} = 20$ & & $\epsilon_{max} = 15$ & \\ \cline{3-12}
& Method & $m(\epsilon)$ & Merit & $m(\epsilon)$ & Merit & $m(\epsilon)$ & Merit & $m(\epsilon)$ & Merit & $m(\epsilon)$ & Merit\\
\hline
\multirow{8}{*}{\rotatebox{90}{\mbox{Semicircle ($n = 102$)}}}
& DPSO   & 12.0(58.0) & 52.4 & 14.8(29.7) & 60.5 & 16.4(23.4) & 60.0 & 18.4(19.8) & 60.2 & 20.0(15.0) & 65.5 \\
& HDPSO1 & 10.0(38.9) & 71.9 & 12.4(29.1) & 80.9 & 13.3(24.0) & 81.0 & 14.8(17.9) & 81.5 & 15.8(13.5) & 92.0 \\
& HDPSO2 & 10.1(58.5) & 70.8 & 12.0(29.5) & 86.7 & 13.0(22.4) & 82.9 & 14.2(19.8) & 85.9 & 15.2(14.4) & 97.0 \\
& ACS    & 10.0(58.6) & 59.0 & 12.6(27.9) & 80.3 & 13.4(23.6) & 81.3 & 16.4(19.9) & 72.6 & 18.0(14.1) & 81.9 \\
& BPS    & 20.0(16.6) & 62.2 & 20.0(16.6) & 62.2 & 20.0(16.6) & 62.2 & 20.0(16.6) & 62.2 & 24.0(8.5) & 73.2 \\
& DPD    & 10.0(38.9) & 100.0 & 12.0(27.3) & 96.7 & 13.0(21.4) & 97.9 & 14.0(18.0) & 97.3 & 15.0(14.4) & 100.0 \\
& AO		 & 12.0(45.3) & 65.7 & 15.0(20.3) & 78.9 & 15.0(20.3) & 78.9 & 19.0(18.3) & 63.0 & 21.0(13.2) & 68.2 \\
& Betweenness&10.0(38.9) & 100.0 & 12.0(28.9) & 93.0 & 13.0(24.8) & 88.6 & 14.0(19.8) & 91.2 & 15.0(14.3) & 100.0 \\
& SP     & 10.0(43.69)& 85.5 & 12.0(26.38)& 98.8 & 13.0(21.41)& 100.0 & 14.0(18.46)& 97.2 & 17.0(14.97)& 87.0\\
& HSP    & 10.0(43.69)& 85.5 & 12.0(26.38)& 98.8 & 13.0(21.41)& 100.0 & 14.0(18.46)& 97.2 & 15.0(14.40)& 100.0\\
		\hline		
		\end{tabular}
\caption{Comparison with other methods using the three benchmark curves.}
\label{tab:resRCPS}
\end{table*}

Considering the same error tolerance $\epsilon_{max}$, results show that the DPSO approach produces an approximating polygon with a relatively larger number of vertices than other approaches. On the other hand, the two hybrid DPSO versions, namely the HDPSO1 and the HDPSO2, perform better than the DPSO, both in terms of the number of polygon vertices and merit. The performance of HDPSO2 is better than that of HDPSO1, as the HDPSO2 uses a more sophisticated optimizer \cite{peng2004}. Our approach has proven to be efficient to compute polygonal approximations with smaller number of vertices for different error $\epsilon_{max}$. Moreover, our results are better than methods such as the ACS in both numbers of vertices and merit.

The BPS method presents a poor performance as it does not consider different levels of details in the curve, i.e., the algorithm computes the redundant points by considering the entire curve information. This reduces the efficiency of the algorithm. The performance of AO method is extremely dependent on the objective function. For the experiments, we considered as the the approximation error $\epsilon$ as the objective function, as in the experiments we always aimed to find a polygonal approximation that respects the maximum approximation error $\epsilon_{max}$.

We notice that our approach performed similarly to the DPD method. For some test cases, the DPD method presented a smaller approximation error $\epsilon$, thus leading to a higher merit (as for the semicircle contour). However, for the leaf contour, our approach presented a smaller number of points, thus leading to a higher merit value. For the same number of points $m$, we notice a smaller approximation error for the DPD method. Such result occurs as DPD method uses a more sophisticated local optimization.

\section{Conclusion}
\label{sec:conclusion}

In this paper, we proposed a novel polygonal approximation method using the networks shortest path. The method represents a contour as a Small-World network, where we represent the points of the curve as network vertices. We represent the network edges by the line segments that compose a possible polygonal approximation of the curve. Then, we compute the contour polygonal approximation as the shortest cycle that exists in this network.

The experiments illustrated that parameter $\epsilon_{1}$ acts as a visibility control in the network vertices, which limits the number of vertices that another vertex can reach to compose an approximation of the arc. We also observed that the initial vertex of the cycle does not interfere in the results, a problem that exists in sequential methods. The method is also deterministic, once it presents the same result over different runs. We also demonstrated that the proposed method is an excellent solution for the problem of polygonal approximation, performing better than traditional and more sophisticated methods. An important feature of the proposed method is its flexibility, as the visibility of $\epsilon_{1}$ enables us to compute different approximations for the same curve. Concerning the Complex Network theory, this work demonstrates a good potential of application of this approach in problems of computer vision and digital imaging processing.

\section{acknowledgements}
O.M.B. gratefully acknowledges the financial support of CNPq (National Council for Scientific and Technological Development, Brazil) (Grant \#308449/2010-0 and \#473893/2010-0) and FAPESP (The State of S\~ao Paulo Research Foundation) (Grant \# 2011/01523-1).

\end{document}